\documentclass[sigconf]{acmart}

\AtBeginDocument{%
  \providecommand\BibTeX{{%
    \normalfont B\kern-0.5em{\scshape i\kern-0.25em b}\kern-0.8em\TeX}}}

\begin{document}

\title{Extensible Database Simulator for Fast Prototyping In-Database Algorithms}

\author{Yifan Wang}
\affiliation{%
  \institution{University of Florida}
  \country{United States}
}
\email{wangyifan@ufl.edu}

\author{Daisy Zhe Wang}
\affiliation{%
  \institution{University of Florida}
  \country{United States}
}
\email{daisyw@ufl.edu}

\begin{abstract}
With the rapid increasing of data scale, in-database analytics and learning has become one of the most studied topics in data science community, because of its significance on reducing the gap between the management and the analytics of data. By extending the capability of database on analytics and learning, data scientists can save much time on exchanging data between databases and external analytic tools. For this goal, researchers are attempting to integrate more data science algorithms into database. However, implementing the algorithms in mainstream databases is super time-consuming, especially when it is necessary to have a deep dive into the database kernels. Thus there are demands for an easy-to-extend database simulator to help fast prototype and verify the in-database algorithms before implementing them in real databases.   

In this demo, we present such an extensible relational database simulator, \textbf{DBSim}, to help data scientists prototype their in-database analytics and learning algorithms and verify the effectiveness of their ideas with minimal cost. DBSim simulates a real relational database by integrating all the major components in mainstream RDBMS, including SQL parser, relational operators, query optimizer, etc. In addition, DBSim provides various interfaces for users to flexibly plug their custom extension modules into any of the major components, without modifying the kernel. By those interfaces, DBSim supports easy extensions on SQL syntax, relational operators, query optimizer rules and cost models, and physical plan execution. To enable accurate evaluation on users' extensions, DBSim supports connecting with real RDBMS and using their real-world cost estimators to calculate the query plan cost. Furthermore, DBSim provides utilities to facilitate users' developing and debugging, like query plan visualizer and interactive analyzer on optimization rules. We develop DBSim using pure Python to support seamless implementation of most data science algorithms into it, since many of them are written in Python.     
\end{abstract}



\maketitle

\section{Introduction}
Though data science is often bound with database, there is still a gap between data science algorithms and databases. Specifically, in the implementations of many data science algorithms, database systems are used only for storing the data, while data processing and analytics are mainly completed outside database. Therefore, data scientists have to spend much time on repeating the loop: exporting data from database, processing the data and doing analytics, re-importing the processed data into database. Today with the super rapid increasing of data scale, this loop is becoming more and more time-consuming.

As a result, the data science and database communities have seen an emergence of the studies about in-database analytics and learning, which aims at making relational databases (RDBMS) natively support general data analytics and some specific data science algorithms. One major research area is in-database machine learning. Many prior works in this area \cite{madlib, indb-distributed-ml-sandha2019database, indb-ml-gpu-schule2021database, indb-ml-db4ml-jasny2020, indb-ml-Declarative-recursive-2019, indb-ml-Learning-Models, indb-ml-Structure-Aware, indb-ml-Vertica-ML, indb-ml-sparse-tensors-abo2018database} extend RDBMS architecture to integrate machine learning algorithms into database, by which users are enabled to call machine learning methods using SQL queries inside the database. Another major area is the extension over relational algebra to introduce new operators and optimization into database, e.g., similarity-aware relational algebra~\cite{extended-algebra-silva2010similarity, extended-algebra-silva2013similarity}, graph-relational algebra~\cite{GRFusion-Hassan2018ExtendingIR}, relational algebra with entity resolution operators~\cite{indb-er-QuERy-2022, indb-er-on-demand-simonini2022entity}, etc.

However, the current mainstream relational databases normally provide no or only limited extensibility to external users, so the main solution to implement new in-database features is modifying the database kernels, which is significantly time-consuming due to the scale of the codebases. Therefore, data scientists demand for a testbed to fast develop the in-database algorithm prototypes and verify whether they work well before starting the implementation in real database systems, which can effectively reduce the waste of time. Thus the testbed should satisfy these requirements: (1) including all the main components of a general RDBMS, (2) being with high extensibility and flexibility, (3) providing debugging and analyzing tools and easy-to-use APIs, and (4) minimal learning cost for users. Unfortunately, no existing tools satisfy all these requirements. Some modular relational query optimizers like Apache Calcite~\cite{calcite-begoli2018apache} and Orca~\cite{orca-soliman2014} have good extensibility for custom optimization rules and cost models, while their flexibility on extending relational operators and query syntax is relatively limited. In addition, most of them are developed in high-performance languages like C++ or Java. Given that many data science algorithms are originally implemented in Python, the in-database algorithms developed with those query optimizers can not be fairly compared to the original versions in terms of performance due to the language difference.

Therefore, we develop DBSim, a pure Python based relational database simulator satisfying those requirements for data science community. To simulate real RDBMS environment, DBSim implements critical database components ranging from SQL parser to query executor, and all of them are with high extensibility and flexibility. DBSim provides APIs for users to extend any of the components, including but not limited to adding new keywords to query syntax, implementing custom operators, writing custom optimization rules, extending physical plan executors, etc. To help users evaluate the efficiency and effectiveness of their extensions, DBSim provides three ways to estimate the query processing cost: (1) using DBSim built-in cost estimator, (2) using the cost estimator of a real DBMS, and (3) measuring cost by the actual query execution time in DBSim. Specifically, DBSim integrates a build-in cost estimator, and supports a more accurate cost estimation by calling the estimator in a real DBMS (e.g., SQLite). Users can also compare the actual query execution time in DBSim with and without their extensions to have the most straightforward insight about the performance difference. 
In addition, a GUI is integrated with DBSim for users to visually and interactively analyze their algorithm prototypes. Furthermore, since DBSim is implemented in pure Python, most data scientists can learn it more easily than C++ or Java tools. We believe this tool can facilitate data science research and benefit data scientists on better understanding the internal of database systems.

In this demonstration, we first introduce the overall architecture, workflow, extension mechanism, the kernel, performance evaluation, APIs and GUI of DBSim in Section~\ref{sec:system-des}. Then we demonstrate how to use DBSim by an example of implementing and optimizing in-database similarity search in Section~\ref{sec:demo}.

\section{System Description}
\label{sec:system-des}
\subsection{Architecture and workflow}
\begin{figure}[!ht]
  \centering
  \includegraphics[width=.9\columnwidth]{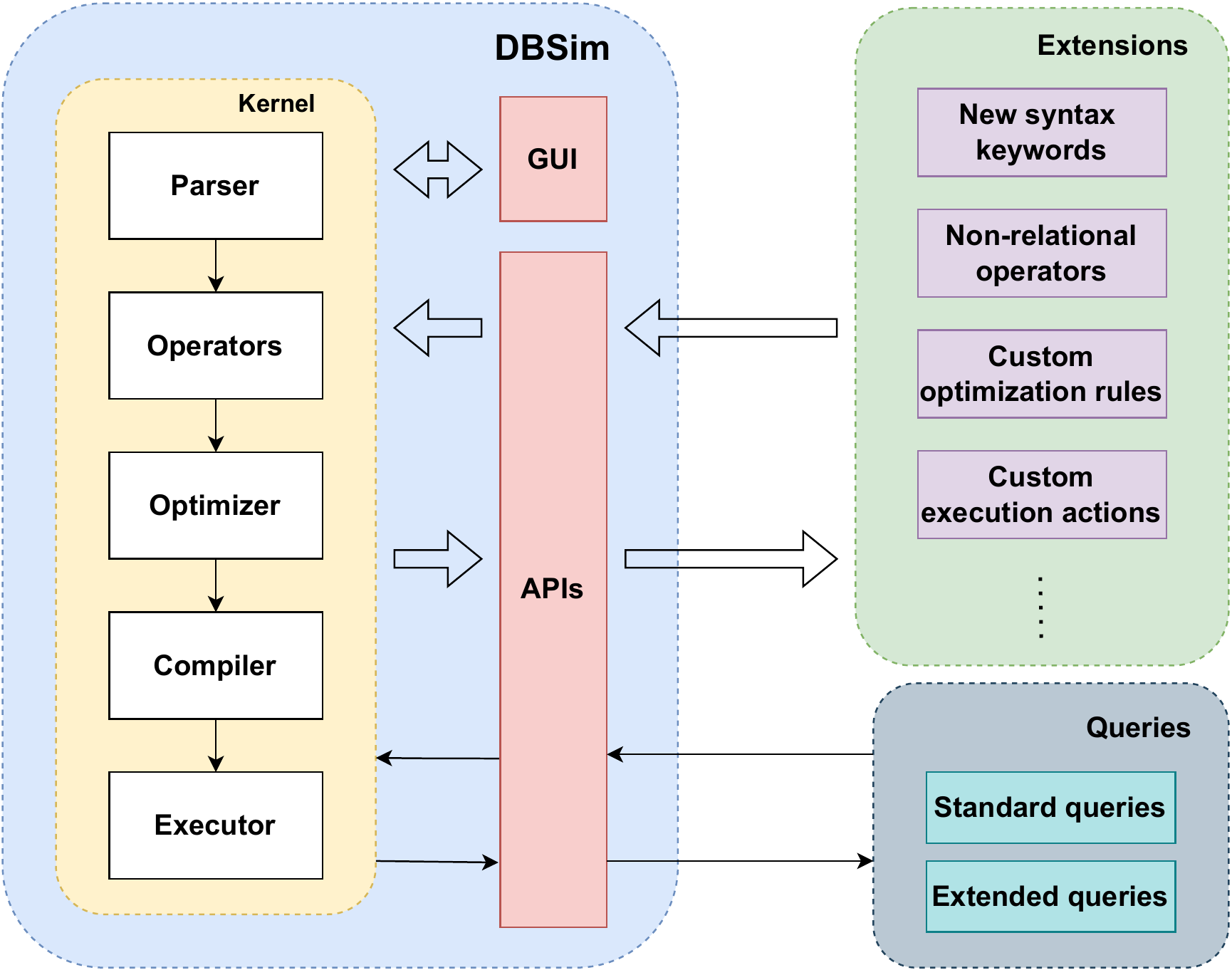}
  \caption{DBSim architecture}
  \setlength{\belowcaptionskip}{-50pt}
  \label{fig:arch}
\end{figure}

The architecture of DBSim is presented in Figure~\ref{fig:arch}. It consists of two parts, the kernel (yellow box) and interfaces (pink boxes). The kernel includes several critical database components (white boxes) from query parser to physical plan executor, while interfaces are the GUI (for interactive querying and analyzing) and APIs (for programmatic querying and building extensions).

There are two workflows in DBSim, one is querying workflow and another is extension workflow, where the former is the steps of query processing and the latter shows how to create an extension and how the extension works in DBSim. We describe more details about them in Section~\ref{sec:kernel} and ~\ref{sec:ext-mech}.

\subsection{Kernel}
\label{sec:kernel}
Similar to the query engine in general RDBMS, the kernel of DBSim includes five components, query parser, operators, query optimizer, query compiler and physical plan executor. Though a real, complete RDBMS normally has more modules like storage engine, it is not necessary to integrate them into DBSim, because most studies about in-database analytics in data science community focus on the query engine. The components of DBSim kernel serve as follows during query processing: (1) query parser parses the input query and generates an abstract syntax tree(AST), (2) operators module resolves the schema information for each node in AST and transforms AST into logical plan, (3) query optimizer applies a series of rules to transform the query plan in order to find the best equivalent plan, (4) query compiler transforms the logical plan into physical plan, and finally (5) physical plan executor executes the plan and outputs results. Based on them, the query processing workflow is same as general RDBMS: as shown in Figure~\ref{fig:arch}, the queries (grey box) are input to the query engine kernel through APIs, then they are passed through the kernel components one by one, from query parser to executor, and finally the executor returns the results to users via APIs. 
Note that we currently implement a heuristic query optimizer (which is more rule-based) instead of the most commonly used volcano-style optimizer (which is more cost-based) in mainstream databases, because rule-based optimizer allows users to fully control the optimization process like the order for applying the rules, which gives users a much clearer insight about how each rule contributes to the optimization. This is important in analyzing the performance and finding bottlenecks of in-database algorithms.    

\subsection{Extension mechanism}
\label{sec:ext-mech}
\begin{figure}[!ht]
  \centering
  \includegraphics[width=.9\columnwidth]{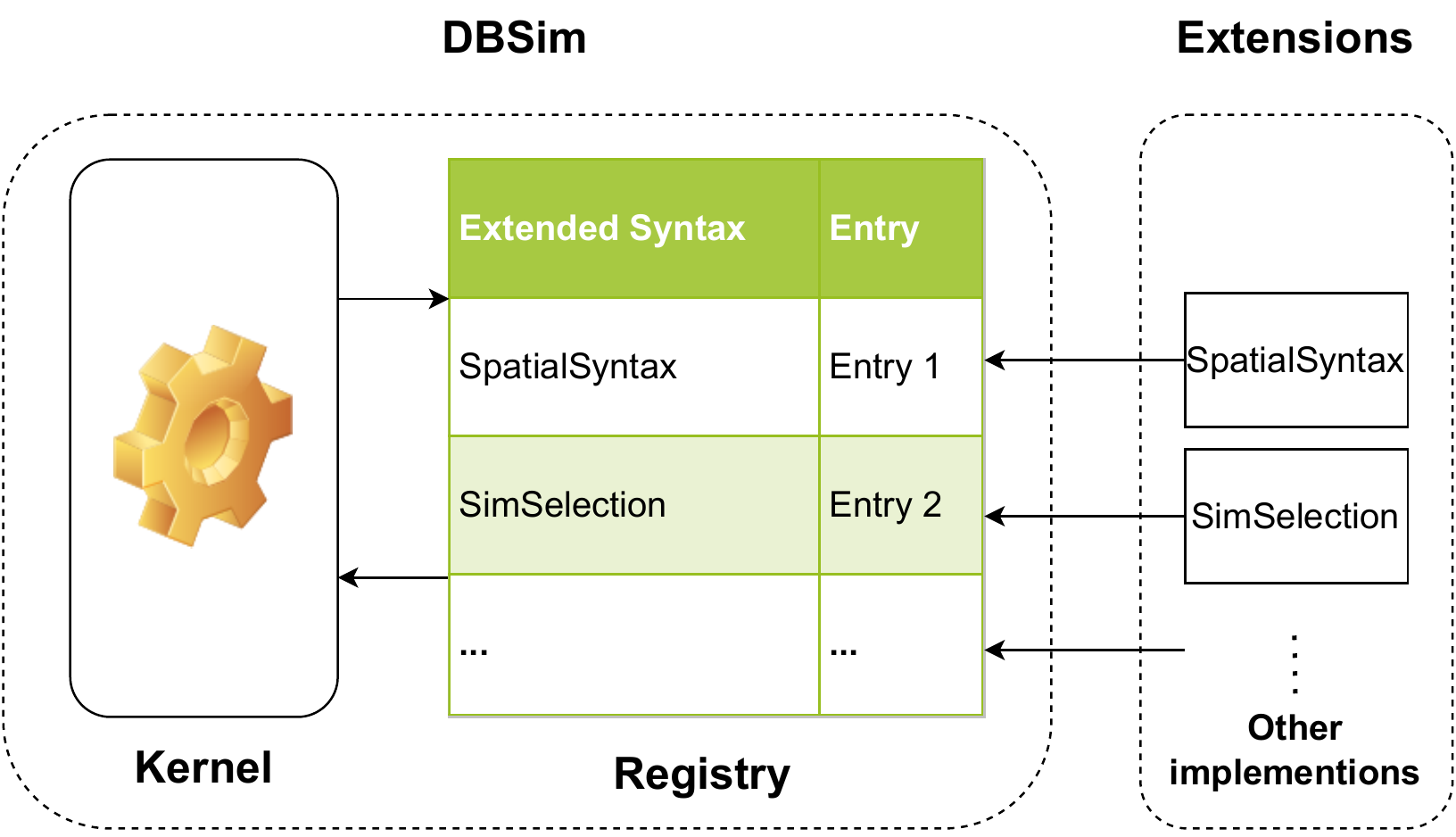}
  \caption{Registry-based extension mechanism}
  \setlength{\belowcaptionskip}{-50pt}
  \label{fig:ext}
\end{figure}

In DBSim, the basic unit for organizing the implementations is \textit{Syntax}, which is a larger concept than the query language syntax. It refers to a scope or name space that manages a collection of implementations ranging from query syntax parsing to query execution functions. We call the native implementations of DBSim \textit{Standard Syntax}. Similarly, the unit of organizing extensions is also Syntax, named as \textit{Extended Syntax}, which is a suite of the implementations for all the extended items scoped in the Syntax. Comparing to Standard Syntax, Extended Syntax is incremental, i.e., it only includes the extended implementations based on Standard Syntax, like adding new operators or overwriting existing functions, instead of re-implementing everything in Standard Syntax.

As one of the most important features of DBSim, flexible extension is supported by a registry-based extension mechanism. As illustrated in Figure~\ref{fig:ext}, the mechanism is centered at a registry that acts as a middleware between the kernel and extensions. Each row in the registry includes the identity of an Extended Syntax and a registry entry which maintains the interfaces for the Syntax to mount its implementations. Specifically, a registry entry provides several entry points for the implementations of different purposes, including adding query keywords, adding data types and operators, inserting optimization rules, customizing the translation from logical to physical plan, etc. To make an Extended Syntax be in effect, users just need to register the implementations of the Syntax to the corresponding entry points in a registry entry and fill the Syntax name and the entry into the registry as a new row. For each registered Extended Syntax, the registry offers an option to enable/disable it, and each registry entry also provides options to enable/disable each entry point. When DBSim starts, the kernel will read in the registry and use the enabled extended implementations to replace the corresponding parts of the Standard Syntax, essentially by the flexible runtime function overwriting in Python.         

The Syntax suite and registry-based mechanism guarantee flexible creation and management of the extensions. Users can enable/disable a whole Extended Syntax simply by turning it on/off in the registry, or manage the extensions in a finer level by manipulating each entry point individually in the registry entry. Such features of DBSim make it quite flexible to apply any one or combination of more than one Extended Syntax (as long as they have no conflict), which has not been achieved by prior works. Figure~\ref{fig:ext} shows an example: user creates two Extended Syntax, namely ``SpatialSyntax'' and ``SimSelection'', to support spatial and similarity queries respectively. The user can freely disable either Syntax or enable both, or disable them partially and let the rests of them co-exist.

\subsection{Performance evaluation}
DBSim provides three methods of query processing cost estimation for users to evaluate the algorithm performance. First, DBSim integrates a built-in cost estimator that calculates the query cost based on multiple factors including data types, operator types,  number of input rows, etc. Second, to make the cost estimation more accurate, DBSim also supports connecting and using the cost estimator of real DBMS which is well-developed and verified in real-world application. Specifically, DBSim translates its query plan and related tables to those of the real DBMS, then feeds them to the DBMS. The translated plan is not exactly executable in the real DBMS, but is normally a valid input for only calling the cost estimator module in the DBMS. Currently DBSim supports SQLite, and the support for PostgreSQL is under development. Third, users can evaluate the performance improvement by comparing the actual query execution time in DBSim with and without their extensions. 

\subsection{Interfaces}
\label{sec:interfaces}

\begin{figure}[!ht]
  \centering
  \includegraphics[width=0.9\columnwidth]{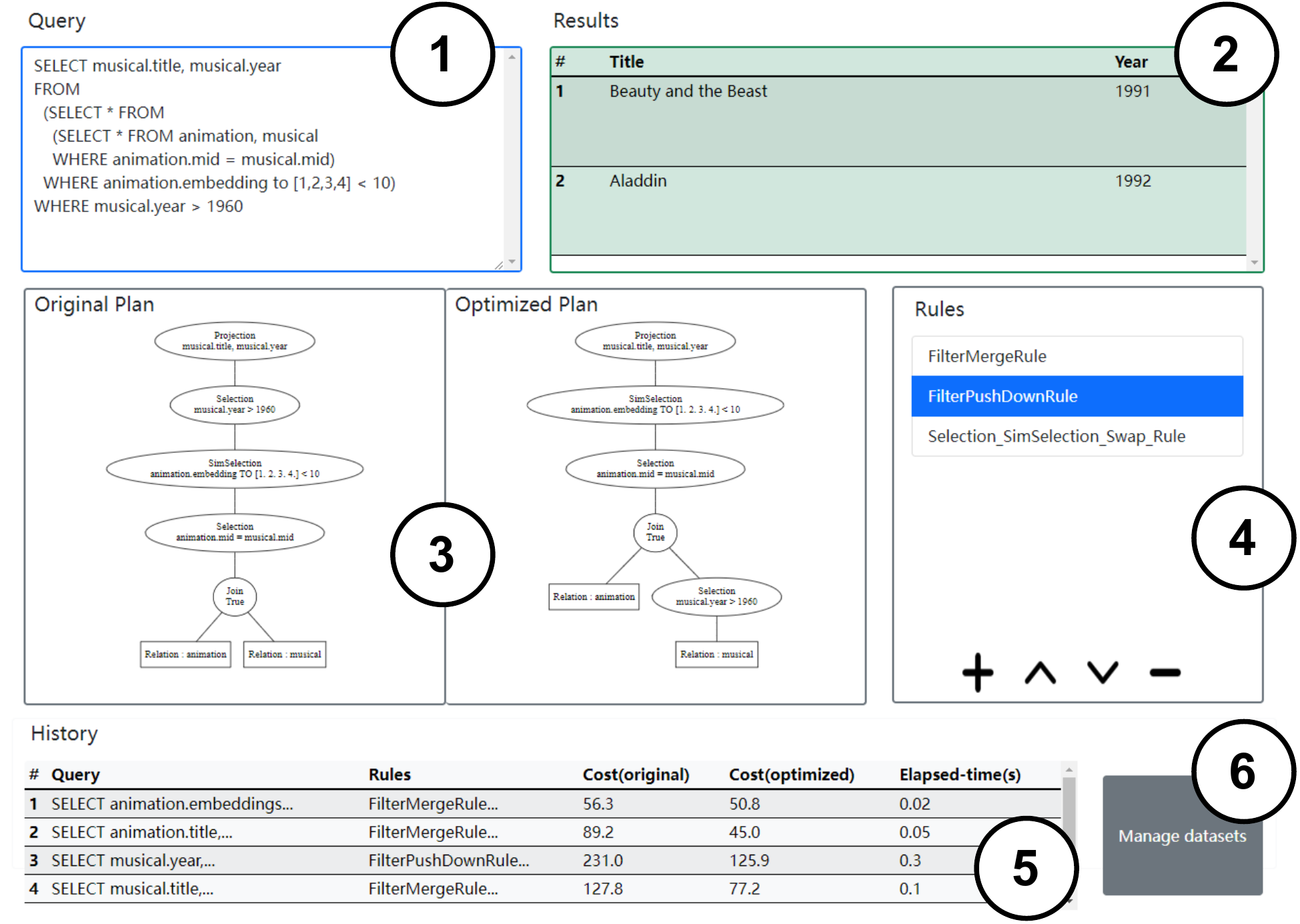}
  \caption{GUI of DBSim}
  \setlength{\belowcaptionskip}{-50pt}
  \label{fig:gui}
\end{figure}

There are two types of interfaces, APIs and GUI, for programmatic and visual access to DBSim respectively. Major APIs are the registry, registry entries and entry points in the entries as discussed in Section~\ref{sec:ext-mech}. In this section we mainly introduce the GUI.

Shown by Figure~\ref{fig:gui}, the GUI includes 6 blocks: (1) query input box, (2) table for showing query execution results , (3) query plan visualization panel, (4) interactive optimization rule list, (5) history records of the input queries, rules applied, estimated plan costs (original and optimized) and real execution time, and (6) the ``Manage datasets'' button for selecting, uploading and deleting datasets. After some datasets are uploaded, when users input a query, it will be automatically executed, with the results reported in block 2 and query plans visualized in panel 3. Panel 3 is spilt into two parts, where the left part presents the initial plan generated directly from the input query and the right part shows the best plan after optimization using the rules in block 4. Block 4 is a menu listing optimization rules being used currently. Users can add/remove rules by clicking the ``+'' or ``-'' buttons, they can also change the order of rules by clicking one of the two arrow signs to move a rule up or down. The query will be optimized by repeatedly applying the listed rules in the given order until no more rules can be applied or a maximum number of iterations is reached. Once a query is executed successfully, the query itself, the rules applied on it, the query cost estimation of the original and optimized plans and the real elapsed time for the execution by DBSim will be recorded and presented in block 5, along with all the previous records. The estimated query cost can be calculated by either DBSim built-in estimator or the real DBMS estimator, depending on the configuration.

With this GUI, users can easily test and analyze whether the Extended Syntax works as their expectation. For example, by the query plan visualization panel, users can know if an extended operator was parsed correctly, and by interactively changing the rules and the order in block 4 and comparing the current plan cost/execution time with the history records, users will acquire a straightforward insight about how the extended rules contribute to the optimization, etc. Furthermore, the GUI can also be used as a teaching tool for users to learn the internal of general RDBMS.           

\section{Demonstration}
\label{sec:demo}
We split the demonstration of DBSim into two phases. In the first phase, we present the steps to implement an Extended Syntax that enables similarity search in relational databases. Then in the second phase, we guide the audience to experience the GUI based on this Extended Syntax. In this section, we first introduce the demonstration use case and datasets we plan to use, and then describe the two demo phases mentioned above.
\subsection{Use case and datasets}
\label{sec:usecase-ds}
We will present this demo based on a use case of in-database movie recommendation. Supposing data scientists are implementing an extended relational algebra that supports in-database similarity search with multi-dimensional embeddings. Based on it they will build an in-database movie recommendation application to recommend movies to users which are similar to their favorites and satisfy some boolean conditions. For example, given an embedding of a user's favorite movie and the fact that the user often watched musical animation movies, the application can use an extended SQL query to find all those movies whose genres include both of ``Musical'' and ``Animation'' and with an embedding similar enough to the given embedding. We will use the MovieLens 20M dataset\footnote{available at https://grouplens.org/datasets/movielens/20m/} 
in this demo. And the embeddings will be generated by some state-of-the-art neural model over the dataset. 

\subsection{Implementing the Syntax}
This is the first phase of this demo. We first introduce the example scenario to the audience as follows: similarity search on vector data is a very commonly used operation in data science algorithms. But it is not supported natively by RDBMS. To implement an in-database similarity search, normally data scientists use user-defined functions (UDF). The problem is that the flexibility of UDF is very limited and its performance is not optimal since query optimizer does not pay as much attention as regular operators to it. Therefore, a better way is extending the relational algebra to add a vector data type and a similarity selection operator which can be seamlessly mixed and optimized with other relational operators.

After the scenario introduction, we present the major steps to implement an Extended Syntax (named as ``SimSelection Syntax'') with the new data type ``Vector'', the new similarity selection operator ``SimSelect'', the new distance computing operator ``To'', and some new optimization rules considering the operators together with other relational operators. The first step is implementing the necessary functions from new query syntax parsing to new execution behaviours. To avoid waste of time, we make the functions ready before this demo and browse them together with the audience at the demo time. If the audience is interested in some details, we will make a further explanation to them, like how the functions parse such a literal ``[1,2,3,4]'' into a Vector. 
The second step is registering the SimSelection Syntax into the registry. We present this step by writing the code at demo time, because it is one of the most important features of DBSim and only a few lines of code need to be written. We show how to create a registry entry, how to mount the functions implemented in the first step onto the entry points, how to enable/disable the whole Syntax or an individual entry point, etc. Finally we load data, execute example queries (with both the extended and standard relational operators, e.g., \textit{SIMSELECT * FROM t1, t2 WHERE t1.c=t2.c AND t1.v TO [1,2,3,4] < 10}) end-to-end through the APIs and show the correct results printed on terminal. 

\subsection{Demonstrating the GUI}
In the first phase of this demo, we present to the audience about how to implement an Extended Syntax, the SimSelection Syntax, in DBSim. Then here in the second phase, we guide the audience to experience the GUI with the SimSelection Syntax. In this section all the block numbers refer to the numbers shown in Figure~\ref{fig:gui}.

First the audience click the ``Manage datasets'' button to select and upload the example datasets we prepare. Then they input a query that mixes up similarity search and regular relational operators into the input box (block 1) and after a while the query will be executed automatically. The audience see the results are printed as a table in result output box (block 2), the initial and optimized query plan trees are visualized in the query plan panel (block 3), and one new record is appended in the history list (block 5), including the current query, the applied rules, the  estimated query costs and the real execution time. 
Then we guide the audience to see the rule menu (block 4). Initially it is blank, therefore the two plans drawn in plan panel are identical. We let the audience click the ``+'' button to browse and select some rules to add into the optimizer. Particularly, we will suggest they to add both of the standard rules and extended rules to see how the extended rules contribute to the optimization. Once it is done, the audience observe that the panel is refreshed, the optimized plan changes and is re-executed, and a new history record is generated accordingly showing a probably lower optimized plan cost and shorter execution time. Then we let the audience try to remove some rules and change the order, and observe how the plan, cost and time change to better understand the optimization process.

\section{Conclusion and future work}
In this demo, we present DBSim, a highly extensible and flexible relational database simulator, for data scientists to quickly prototype in-database analytics and learning algorithms, such that they can verify whether their ideas work well with a minimal cost before implementing the ideas in real RDBMS. It provides enough extensibility and flexibility for users to easily implement in-database versions of many data science algorithms. Its easy-to-use APIs and interactive GUI facilitate users in testing and analyzing their custom extensions. Furthermore, being implemented in pure Python significantly reduces its learning cost for data scientists. As future work, we plan to implement index and volcano-style query optimizer in DBSim, as well as support more real-world RDBMS for query cost estimation.

\begin{acks}
This work is partially supported by DARPA under Award \#FA8750-18-2-0014 (AIDA/GAIA). 
\end{acks}

\bibliographystyle{ACM-Reference-Format}
\bibliography{main}

\end{document}